\documentclass[12pt]{article}
\usepackage{epsfig}

\begin{document}
\mbox{ }\\[-1cm]
\mbox{ }\hfill \\
\mbox{ }\hfill \\

\vskip 1cm

\begin{center}
{\Large\bf Transverse Beam Dynamics including Aberration
           Effects in the Thermal Wave Model
           using a Functional Method}\\[1cm]   
Ji-ho Jang\footnote{jangjh@kaeri.re.kr}, Yong-sub Cho, Hyeok-jung Kwon
\end{center}

\bigskip 

\begin{center}
{\it Korea Atomic Energy Research Institute, Daejeon 305-353, Korea}
\end{center}

\vskip 2.5cm
We studied the transverse beam dynamics including aberration effects of
sextupole and octupole perturbations in a thermal wave model. 
A functional integration method was
used to calculate the first-order perturbation effects. We found that the
model successfully explains a PARMILA simulation results for proton beams 
without space-charge effects in a FODO lattice.
\begin{abstract}

\vskip 0.5cm

\end{abstract}
\newpage

The thermal wave model (TWM) can be used to study the beam dynamics for 
relativistic charged particles \cite{FM:1991}. 
This model was used to estimate the luminosity for a beam perturbed by the 
spherical aberration in a linear collider \cite{FM:1992} and 
explain the halo formation by using a Gaussian slit \cite{KP:2000}. 
The transverse beam dynamics of charged particles 
is a promising branch of the TWM. In ref. \cite{FGM:1994}, 
authors showed that the particle 
distribution was successfully described by the TWM in a thin lens approximation 
of a quadrupole magnet with aberration.
It can be extended to a general quadrupole lattice for a Gaussian beam by
using a functional integration method if the space charge effects can be
neglected \cite{JCK:2007}. 
However the work was restricted to a perfect quadrupole magnet. 
In this paper, we extended the TWM formulation for a transverse beam dynamics 
to include a sextupole and an octupole perturbations 
by using the first-order perturbation theory of a functional method. 
The model calculation was compared to the PARMILA \cite{TB:2002} 
simulation result in a FODO lattice. 
We found that the model can explain the PARMILA results successfully if the
perturbation effect is not too large.

In TWM, the beam wave function of the relativistic charged
particle can be described by the Schr\"{o}dinger-type equation as following,
\begin{eqnarray}
\label{eq:differential}
i~\epsilon~\frac{\partial~\psi(x,z)}{\partial z} =
-~\frac{\epsilon^2}{2}~\frac{\partial^2}{\partial x^2}~\psi(x,z)
+ U(x,z)~\psi(x,z),
\end{eqnarray}
where $z = ct$ is the longitudinal coordinate of the beam motion and 
$U(x,z) \equiv u(x,z)/m_0 \gamma \beta^2 c^2$ is the dimensionless potential.
The transverse beam distribution function is given by a modulus squared of the 
beam wave function, 
$ \rho(x,z) = N \left| \psi(x,z) \right|^2$
with a particle number of $N$.
The beam wave function satisfies the following normalization condition,
$\int_{-\infty}^{\infty} \left| \psi(x,z) \right|^2 dx = 1$.

The initial Gaussian beam wave function can be described by
\begin{eqnarray}
\psi_1(x,0)=\left( \frac{1}{2 \pi \sigma_1^2} \right)^\frac{1}{4}
          \exp \left[-\frac{x^2}{4 \sigma_1^2} + 
          i \left( \frac{x^2}{2 \epsilon \rho_1} + \theta_1 \right) \right].
\end{eqnarray}
where $\epsilon = 2 \epsilon_{\rm rms}$ with the unnormalized rms emittance 
of $\epsilon_{\rm rms}$.
The parameters, $\sigma_1$ and $\rho_1$ are related to the twiss parameters
of the initial particle distribution through 
$\sigma_1 = \sqrt{\epsilon_{\rm rms} \beta_1}$ and
$\rho_1 = - \beta_1/\alpha_1$ \cite{JCK:2007}.
An elegant way to solve the differential equation 
(Eq.~(\ref{eq:differential})) is a functional integral method
where the solution is given by the product of kernel (or propagator) and the
initial beam wave function,
\begin{eqnarray}
\label{eq:formalsol}
\psi(x_2,z) =  \int_{-\infty}^{\infty} dx_1 K(x_2,z ; x_1,0)~\psi(x_1,0).
\end{eqnarray}


When we include the sextupole and octupole perturbation, the potential
can be described as
\begin{eqnarray}
V(x) = U_0(x) - U_p(x),
\end{eqnarray}
with
\begin{eqnarray}
U_0(x) = \frac{k_2}{2} x^2, ~~~~
U_p(x) = \frac{k_3}{3 !} x^3 +\frac{k_4}{4 !} x^4.
\end{eqnarray}
where $U_0(x)$ is the potential of a quadrupole magnet and 
$U_p(x)$ describes the aberration effects.
The $k_2,~k_3$, and $k_4$ are related to the field strength of the quadrupole,
sextupole, and octupole magnets. 

In the first-order perturbation,
the kernels for a focusing and defocusing quadruple magnets become
\begin{eqnarray}
K_f(x_2,z \mbox{;}x_1,0 \mbox{;}V_p(x))&=& K^{(0)}_f (x_2,z\mbox{;}x_1,0)
\left( 1+ \sum_{n=0}^4 h_n(x_2,z) \left( \frac{x_1}{\sigma_0} \right)^n 
\right),\\
K_d(x_2,z \mbox{;}x_1,0 \mbox{;}V_p(x))&=& K^{(0)}_d (x_2,z\mbox{;}x_1,0)
\left( 1+ \sum_{n=0}^4 g_n(x_2,z) \left( \frac{x_1}{\sigma_0} \right)^n 
\right).
\end{eqnarray}
where $K^{(0)}_f (x_2,z\mbox{;}x_1,0)$ 
is the unperturbed kernel for a focusing case as given in Ref. \cite{JCK:2007}:
\begin{eqnarray}
K^{(0)}_f (x_2,z;x_1,0) &=& \left( \frac{\sqrt{k_1}} 
   {2 \pi i \epsilon \sin (\sqrt{k_1}z)} \right)^{1/2} 
   e^{i \frac{\sqrt{k_1}}{2 \epsilon} \left[(x_2^2+x_1^2) \cot \sqrt{k_1}z -
   2 x_2 x_1 \csc \sqrt{k_1} z \right]}
\end{eqnarray}
The kernel $K^{(0)}_d$ for the defocusing case is obtained by replacing 
cot and csc functions in
the focusing case with coth and csch functions, respectively.
From the functional perturbation theory \cite{H:1992}, 
we obtained the coefficient
functions for a focusing quadrupole as follows,
\begin{eqnarray}
h_0(x,z) &=& \left[  -s_1(z) \left( \frac{x}{\sigma_0} \right)+
            i \frac{s_3(z)}{12} \left( \frac{x}{\sigma_0} \right)^3 \right] r_3
\nonumber \\
      &+& \left[ -i \frac{3 s_4^2(z) s_5(z)}{8}  
          + \frac{3 s_4^3(z) s_6(z)}{16} \left( \frac{x}{\sigma_0} \right)^2 
          + i \frac{s_4^4(z) s_7(z)}{64} \left( \frac{x}{\sigma_0} \right)^4
          \right] r_4,  \nonumber \\
h_1(x,z) &=&  \left[ -s_1(z)  
                  + i \frac{s_2(z)}{4} \left( \frac{x}{\sigma_0} \right)^2
             \right] r_3
- \left[ \frac{3 s_4^3(z) s_5(z)}{4} \left( \frac{x}{\sigma_0}
                                                \right) 
          + i \frac{s_4^4(z) s_6(z)}{16} \left( \frac{x}{\sigma_0} \right)^3 
          \right] r_4, \nonumber \\
h_2(x,z) &=& i \frac{s_2(z)}{4} \left( \frac{x}{\sigma_0} \right) r_3
  + \left[ \frac{3 s_4^3(z) s_6(z)}{16} 
     +i \frac{3 s_4^4(z) s_5(z)}{16} \left( \frac{x}{\sigma_0} \right)^2 
  \right] r_4, \nonumber \\
h_3(x,z) &=& i \frac{s_3(z)}{12} r_3 
        - i \frac{s_4^4(z) s_6(z)}{16} 
        \left( \frac{x}{\sigma_0} \right) r_4,  \nonumber \\
h_4(x,z) &=&  i \frac{s_4^4(z) s_7(z)}{64} r_4,
\end{eqnarray}
where $r_3 = k_3 \sigma_0/(6 k_1)$ and $r_4 = k_4 \sigma_0^2/(24 k_1)$ with
$\sigma_0^2 = \epsilon /(2 \sqrt{k_1})$.
As noted in Ref. \cite{FGM:1994}, $r_3$ and $r_4$ are the perturbation parameters.
The functions $s_i(z)$ are given by
\begin{eqnarray}
s_1(z) &=& \tan^2(\sqrt{k_1} z/2),\nonumber \\
s_2(z) &=& \sec^2(\sqrt{k_1} z/2) \tan(\sqrt{k_1} z/2),\nonumber \\
s_3(z) &=& (2+\sec^2(\sqrt{k_1} z/2))\tan(\sqrt{k_1} z/2),\nonumber \\
s_4(z) &=& \csc(\sqrt{k_1} z),\nonumber \\
s_5(z) &=& 4 \sqrt{k_1} z+2\sqrt{k_1} z \cos(2\sqrt{k_1} z) 
        - 3\sin(2\sqrt{k_1} z),\nonumber \\
s_6(z) &=& 12 \sqrt{k_1} z\cos(\sqrt{k_1} z)
        -9\sin(\sqrt{k_1} z)-\sin(3\sqrt{k_1} z),\nonumber \\
s_7(z) &=& 12 \sqrt{k_1} z -8\sin(2\sqrt{k_1} z)+\sin(4\sqrt{k_1} z).
\end{eqnarray}
The corresponding functions $g_i(x,z)$ for a defocusing case 
can be obtained by replacing the trigonometric functions in $h_i(x,z)$ 
with the corresponding hyperbolic functions.
For a drift space, the unpertubed kernel is enough to get the final wave
function:
\begin{eqnarray}
K^{(0)}_0 (x_2,z;x_1,0) &=& \left( \frac{1}{2 \pi i \epsilon L}\right)^{1/2}
e^{\frac{i}{2 \epsilon z} (x_2-x_1)^2}.
\end{eqnarray}
where $z$ is the length of the drift space.

Because the beam wave function includes some perturbed terms after passing 
a quadrupole magnet with aberration, we generalize the initial beam wave 
function like
\begin{eqnarray}
\psi_1(x,0)=\left( \frac{1}{2 \pi \sigma_1^2} \right)^\frac{1}{4}
          \exp \left[-\frac{x^2}{4 \sigma_1^2} + 
          i \left( \frac{x^2}{2 \epsilon \rho_1} + \theta_1 \right) \right]
          \left\{ 1+ \sum_{n=0}^4 a_n \left( \frac{x}{\sigma_0}
                                      \right)^n \right\}
\end{eqnarray}
where $a_n = {\cal{O}} (k_3, k_4)$ describes the perturbation effects.

After a drift space, the wave function becomes
\begin{eqnarray}
\psi_2(x,z)
&=& \int dx_1 K_0^{(0)} (x,z ; x_1, 0) \psi_1 (x_1, 0) \nonumber \nonumber \\
&=& \left( \frac{1}{2 \pi \sigma_2^2} \right)^\frac{1}{4}
          \exp \left[-\frac{x^2}{4 \sigma^2_2} + 
          i\left( \frac{x^2}{2 \epsilon \rho_2} +\theta_1+\theta_2\right)\right]
          \left\{ 1+
          \sum_{n=0}^4 
          a_n \left( \frac{\sqrt{2} \sigma_1}{\sigma_0} \right)^n  
          I_n(\frac{x}{\sqrt{2}\sigma_2},\theta_2)
          \right\}, \nonumber \\
&&           
\end{eqnarray}
where the parameters $\theta_2, \sigma_2$, and $\rho_2$ after the drift space
are related to the corresponding initial parameters
$\theta_1, \sigma_1$, and $\rho_1$. 
The explicit formula can be found in Ref. \cite{JCK:2007}.
The functions $I_n (x,\theta)$ in the final wave function are given by
\begin{eqnarray}
I_0(x,\theta) &=& 1,~~
I_1(x,\theta) = \frac{1}{2} e^{2 i \theta} H_1(x),~~
I_2(x,\theta) = \frac{1}{4}  e^{4 i \theta}  H_2(x) +\frac{1}{2}, \nonumber \\
I_3(x,\theta) &=& \frac{1}{8}  e^{6 i \theta} H_3(x)
   + \frac{3}{4}  e^{2 i \theta}  H_1(x), ~~
I_4(x,\theta) =\frac{1}{16}  e^{8 i \theta} H_4(x)
   + \frac{3}{4}  e^{4 i \theta}  H_2(x) 
    +\frac{3}{4},
\end{eqnarray}
where $H_n(x)$ are the Hermite polynomials.

After a focusing lens of the effective length of $z$, the wave function becomes
\begin{eqnarray}
\psi_2(x,z)
&=& \int dx_1 K_f (x,z ; x_1, 0) \psi_1 (x_1, 0) \nonumber \\
&=& \left( \frac{1}{2 \pi \sigma^2_2} \right)^\frac{1}{4}
          \exp \left[-\frac{x^2}{4 \sigma^2_2} + 
          i \left( \frac{x^2}{2 \epsilon \rho_2} + \theta_1 + \theta_2 \right) 
          \right] 
          \left\{ 1 + \sum_{n=0}^4 (h_n + a_n)
          \left( \frac{\sqrt{2} \sigma_1}{\sigma_0} \right)^n 
          I_n(\frac{x}{\sqrt{2}\sigma_2},\theta_2)
          \right\} \nonumber \\
&&          
\end{eqnarray}
For a defocusing case, the function $h_n$ is replaced with $g_n$.
The initial and final parameters, $\theta_i$, $\sigma_i$, and $\rho_i$, 
can be obtained by the same formula as given in Ref. \cite{JCK:2007}.

We note that the relation between initial model parameters, 
$\sigma_1$ and $\rho_1$,
and resulting parameters, $\sigma_2$ and $\rho_2$, can be obtained by a 
well-known transformation of the twiss parameters \cite{W:1998},
\begin{eqnarray}
\left( 
  \begin{array}{c}
  \beta_2 \\ \alpha_2 \\ \gamma_2
  \end{array}
\right) =
\left(   
  \begin{array}{ccc}
  R_{11}^2       &  -2 R_{11}R_{12}  & R_{12}^2       \\
  - R_{11}R_{21} &  1+2 R_{12}R_{21} & - R_{12}R_{22} \\
  R_{21}^2       &  -2 R_{21}R_{22}  & R_{22}^2
  \end{array}
\right)
\left(  
  \begin{array}{c}
  \beta_1 \\ \alpha_1 \\ \gamma_1
  \end{array}
\right)  
\end{eqnarray}
with $\gamma_i \beta_i - \alpha_i^2 = 1$.
The twiss parameters are related to the TWM paramters through 
$\sigma_i = \sqrt{\epsilon_{\rm rms} \beta_i}$ and 
$\rho_i = - \beta_i/\alpha_i$.

We also note that the final wave function is not normalized to one even though
the initial wave function is normalized to be one. 
The reason is that the modulus squared of the wave function includes the 
second-order terms which are proportional to ${\cal{O}} (k_3^2, k_4^2)$. 
Because the particle distribution function becomes negative in some region
if we keep the linear terms only in modulus squared, we have to use a modulus
squared for the particle distribution. 
Hence the normalization increase should be an indicator to evaluate the
validity of the perturbation calculation.
We found that normalization increases less than few percent from one, 
the perturbation calculation described the PARMILA calculation successfully.

In order to check the perturbation calculation, we compared the results with
the PARMILA simulation with 50,000 macro particles which pass through
a FODO lattice. 
The field gradient and the effective length of a quadrupole magnet are
10 T/m and 0.2 m, respectively. The length of a drift space is 0.5 m.
In the simulation, we assumed that the field strengths of the sextupole and 
octupole components are 3\% of the quadrupole strength. 
We used matched input beam of the lattice with the twiss parameters,
$\alpha = -1.60$, $\beta=2.37$ mm/mrad, and 
$\epsilon_{\rm rms} = 1.56$ $\pi$ mm-mrad.
They correspond to $\sigma_1 = 1.92$ mm and $\rho_1 = 1.48$ m.
We used a Gaussian input beam and neglected the space charge effects.

Figure 1 shows the input particle distribution (histogram) for the PARMILA
calculation and the model result (solid line).
The input beam wave function is normalize to one in the mks unit.
The particle distributions after each beam-optical component are given in
from Figure 2 to Figure 5. 
The solid lines in figures show the TWM result
with sextupole and octupole perturbations. 
We also included the pure quadrupole result as dotted lines in the figures
to evaluate the validity of the results with and without aberration effects.
The PARMILA calculation shows clearly that particle distribution deviates 
from a Gaussian distribution. 
The TWM with aberration successfully described the asymmetric result
as shown in the figures.
In order to compare the results quantitatively, 
we calculated $\chi$ values in each step. 
The result is summarized in Table 1. 
It shows that the TWM with aberration improves the result 
than the model without perturbation.

This work is related to the TWM calculation for a transverse beam dynamics
of the charged particle with a sextupole and octupole perturbations.
This model calculation is valid if space charge effects are negligible. 
Even though the perturbation result is limited to the first-order, 
it explained the PARMILA simulation results successfully 
when the aberration field strength is less than few percent of
the quadrupole strength.

\begin{center}
{\bf ACKNOWLEDGEMENTS}
\end{center}

This work was supported by the Ministry of Education,
Science, and Technology of the Korean government.

\vspace{0.5cm}

\newpage
\begin{table}[htb]
\caption{\label{table:chi}
         $\chi$ values between PARMILA result and TWM calucation with
         and without aberration effects.}
\begin{center}
\begin{tabular}{l|ccccc}\hline 
               & initial          & focusing quad.   & 1st dirft 
               & defocusing quad. & 2nd drift \\ \hline
TWM without aberration & 92.1 & 104.2 & 170.2 & 168.5  & 145.0 \\
TWM with aberration    & 92.1 & 105.1 & 111.5 & 102.3  & 108.4 \\ \hline
\end{tabular}
\end{center}
\end{table}

 \begin{figure}[htb]
 \centering
 \hspace{-1cm}
  \includegraphics[angle=0, scale=1.2]{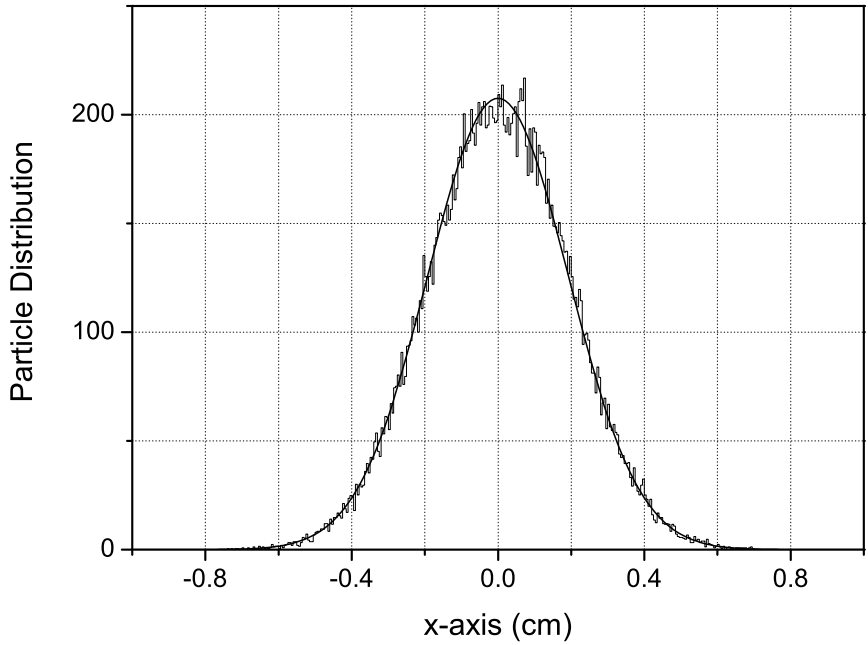}
  \label{fig:input_beam}
 \caption{Particle distributions of the input beam. 
          The histogram and solid line represent the 
          PARMILA results and the model predictions, respectively.}
 \end{figure}
 
 \begin{figure}[htb]
 \centering
 \hspace{-1cm}
  \includegraphics[angle=0, scale=1.2]{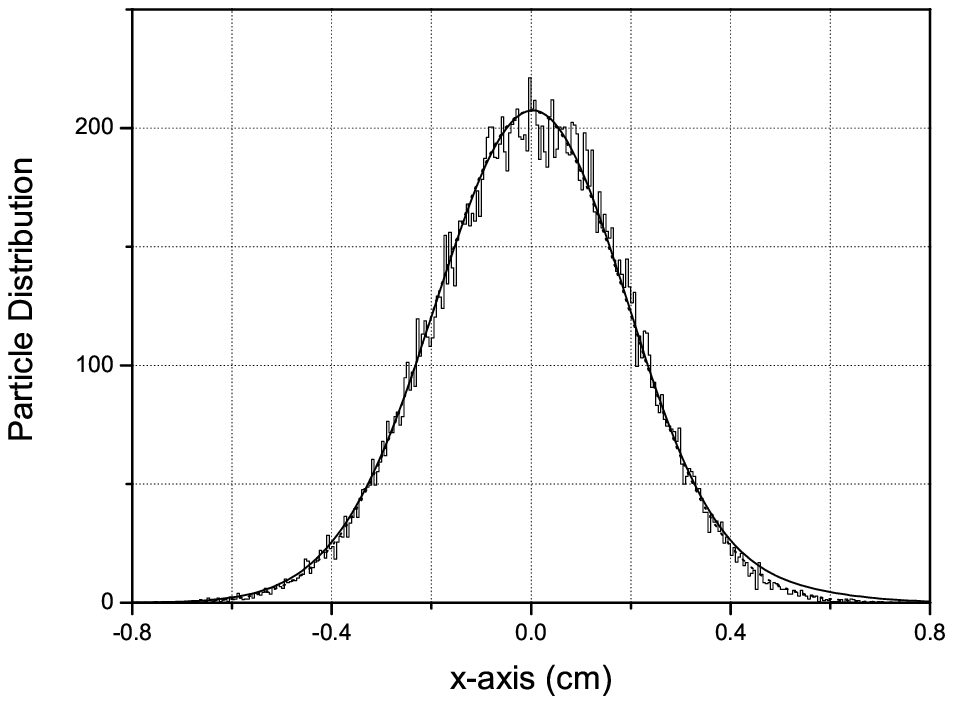}
  \label{fig:beamF}
 \caption{Particle distributions after the first quadrupole magnet. 
          The histogram, solid, and dotted lines represent the 
          PARMILA results and the model predictions with and without 
          aberration, respectively.}
 \end{figure}
 
 \begin{figure}[htb]
 \centering
 \hspace{-1cm}
  \includegraphics[angle=0, scale=1.2]{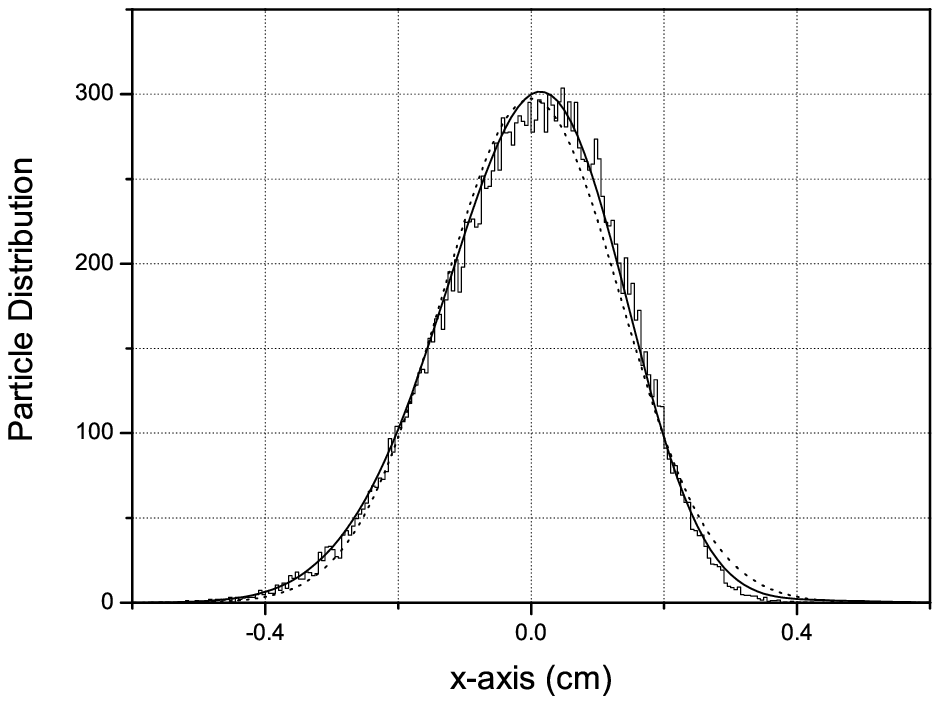}
  \label{fig:beamFO}
 \caption{Particle distributions after the first drift space. 
          The histogram, solid, and dotted lines represent the 
          PARMILA results and the model predictions with and without 
          aberration, respectively.}
 \end{figure}
 
 \begin{figure}[htb]
 \centering
 \hspace{-1cm}
  \includegraphics[angle=0, scale=1.2]{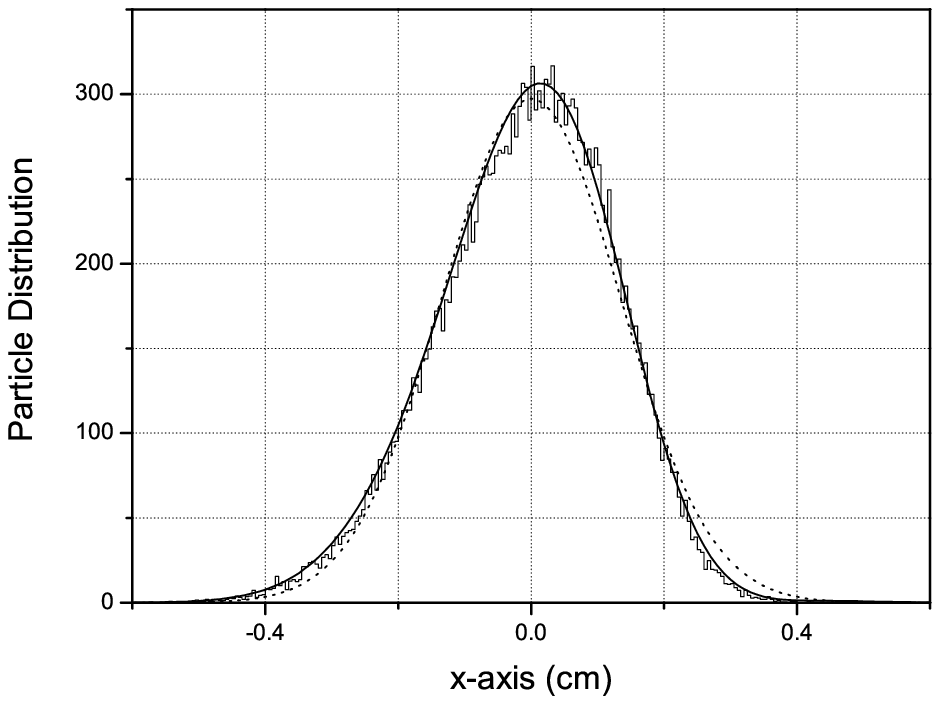}
  \label{fig:beamFOD}
 \caption{Particle distributions after the second quadrupole magnet. 
          The histogram, solid, and dotted lines represent the 
          PARMILA results and the model predictions with and without 
          aberration, respectively.}
 \end{figure}
 
 \begin{figure}[htb]
 \centering
 \hspace{-1cm}
  \includegraphics[angle=0, scale=1.2]{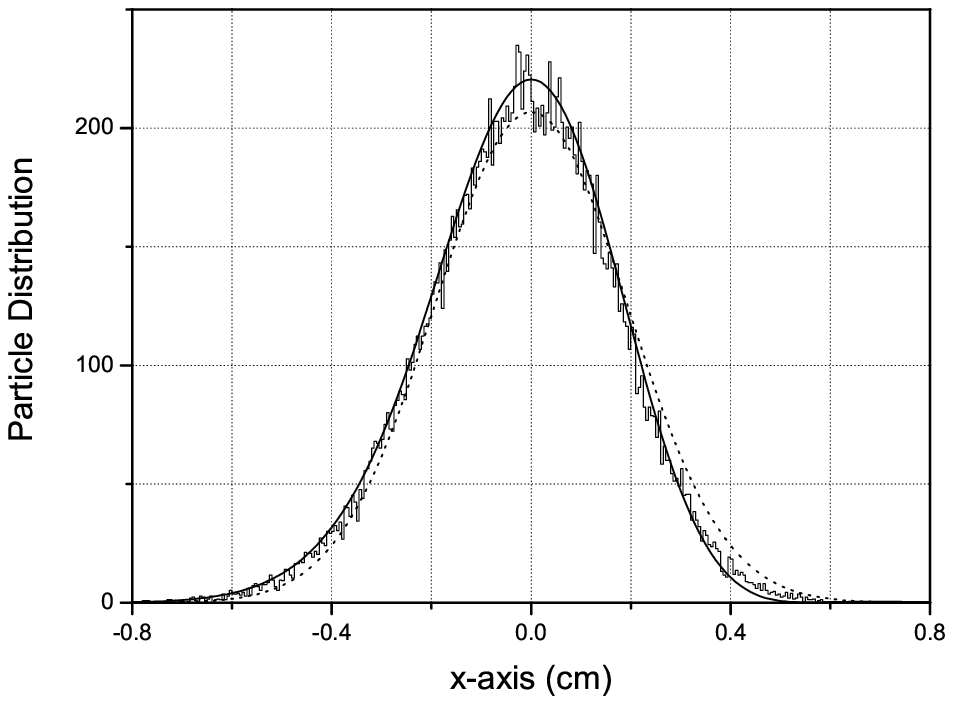}
  \label{fig:beamFODO}
 \caption{Particle distributions after the second drift space. 
          The histogram, solid, and dotted lines represent the 
          PARMILA results and the model predictions with and without 
          aberration, respectively.}
 \end{figure}


\begin{thebibliography}{99}

\bibitem{FM:1991} R. Fedele and G. Miele, Nuovo Cimento D 13 (1991) 1527. 

\bibitem{FM:1992} R. Fedele and G. Miele, Phys. Rev. A 46 (1992) 6634. 

\bibitem{KP:2000} S. A. Khan and M. Pusterla, Eur. Phys. J. A 7 (2000) 583.
                 
\bibitem{FGM:1994} R. Fedele, F. Galluccio and G. Miele, 
                   Phys. Lett. A 185 (1994) 93. 
                   
\bibitem{JCK:2007} J. Jang, Y. Cho and H. Kwon, 
                   Phys. Lett. A 366 (2007) 246.
                   
\bibitem{TB:2002} H. Takeda and J. Billen, PARMILA, LA-UR-98-4478.

\bibitem{H:1992} H. Holstein, Topics in Advanced Quantum Mechanics 
                 (Addison-Wesley, 1992).

\bibitem{W:1998} T. Wangler, Principles of RF Linear Accelerators,
                 (John Wiley \& Sons, 1998)
                                  

\end{thebibliography}
\end{document}